\documentclass[sigconf, nonacm]{acmart}

\newcommand\vldbdoi{XX.XX/XXX.XX}

\newcommand\vldbavailabilityurl{URL_TO_YOUR_ARTIFACTS}
\newcommand\vldbpagestyle{plain} 
\usepackage{setspace}
\usepackage{tikz}
\usepackage{makecell}
\usepackage{amsthm}
\newtheoremstyle{compactthm}
  {4pt}        
  {4pt}        
  {\itshape}   
  {}           
  {\bfseries}  
  {.}          
  {.5em}       
  {}

\theoremstyle{compactthm}
\newtheorem{assumption}{Assumption}
\newtheorem{proposition}{Proposition}

\newcommand{\circledchar}[2][gray!70]{%
    \tikz[baseline=(char.base)]{
        \node[shape=circle,draw=#1,fill=#1,text=white,inner sep=0.25pt,scale=0.75] (char) {#2};
    }
    }

\begin{document}
\title{Topology-Aware LLM-Driven Social Simulation: A Unified Framework for Efficient and Realistic Agent Dynamics}

\author{Yuwei Xu}
\authornote{These authors contributed equally to this work.}
\affiliation{%
  \institution{The Chinese University of \\ Hong Kong, Shenzhen}
  \city{Shenzhen}
  \country{China}
}
\email{yuweixu@link.cuhk.edu.cn}

\author{Shulun Zhang}
\authornotemark[1]
\affiliation{%
    \institution{The Chinese University of \\ Hong Kong, Shenzhen}
    \city{Shenzhen}
    \country{China}
}
\email{shulunzhang@link.cuhk.edu.cn}

\author{Yingli Zhou}
\authornote{Open-source project leader.}
\affiliation{%
    \institution{The Chinese University of \\ Hong Kong, Shenzhen}
    \city{Shenzhen}
    \country{China}
}
\email{yinglizhou@link.cuhk.edu.cn}

\author{Shipei Zeng}
\affiliation{%
  \institution{Shenzhen Research Institute \\ of Big Data}
    \city{Shenzhen}
    \country{China}
}
\email{shipei.zeng@sribd.cn}

\author{Laks V.S. Lakshmanan}
\affiliation{%
  \institution{The University of British Columbia}
  \city{Vancouver}
  \country{Canada}
}
\email{laks@cs.ubc.ca}

\author{Chenhao Ma}
\authornote{Corresponding author.}
\authornotemark[2]
\affiliation{%
    \institution{The Chinese University of \\ Hong Kong, Shenzhen}
    \city{Shenzhen}
    \country{China}
}
\email{machenhao@cuhk.edu.cn}

\newcommand{\method}{TopoSim}

\begin{spacing}{0.99}

\begin{abstract}
LLM-driven social simulation has emerged as a promising paradigm for modeling social-network dynamics by representing users as LLM agents and simulating their interactions over a social graph.
In these simulators, agents repeatedly gather messages from neighbors, update their states through LLM reasoning, and scatter responses back to the network.
As simulations scale to larger social networks, this workflow creates a fundamental scalability challenge: existing LLM-based simulators typically treat the social network as a communication scaffold and execute graph-coupled agent updates independently in each round, requiring costly LLM inference for every agent while repeatedly aggregating similar graph-dependent contexts and maintaining evolving states over time.
As a result, network topology is largely lost as an execution signal, leading to redundant inference over shared social context and obscuring receiver-dependent influence exposure.
To address this, we present \method{}, a topology-aware execution layer that converts network topology into an execution prior for scalable and faithful LLM social simulation.
\method{} realizes this prior by materializing receiver-dependent influence into prioritized neighborhood views and coordinating structurally compatible update requests into shared inference cells.
Experiments across simulation workloads, LLM backbones, graph scales, and real-world trajectories show that \method{} preserves simulation fidelity, improves alignment with observed dynamics, and substantially reduces execution cost.
These results demonstrate that topology can be exploited not only as part of the simulated environment, but also as a practical execution signal for scaling graph-structured LLM simulation.
\end{abstract}

\maketitle

\pagestyle{\vldbpagestyle}

\ifdefempty{\vldbavailabilityurl}{}{
\vspace{.3cm}
\begingroup\small\noindent\raggedright\textbf{Artifact Availability:}\\
The source code, data, and/or other artifacts have been made available at \url{https://github.com/D2I-CUHKSZ/MicroWorld}.
\endgroup
}

\section{Introduction}
In real-world social systems, countless interactions among individuals give rise to complex, emergent macro-level social dynamics. 
However, directly observing or intervening in such systems is often infeasible due to their scale, cost, and privacy constraints. 
Social simulation addresses this challenge by modeling micro-level interactions to study macro-level phenomena such as epidemic spreading and opinion influence, providing an important tool for understanding complex social systems~\cite{Eubank2004, Ferguson2005H5N1, Bakshy2011influencer, Watts2007}.

\begin{figure*}[]
  \begin{center}
    \centerline{\includegraphics[width= 0.90\textwidth]{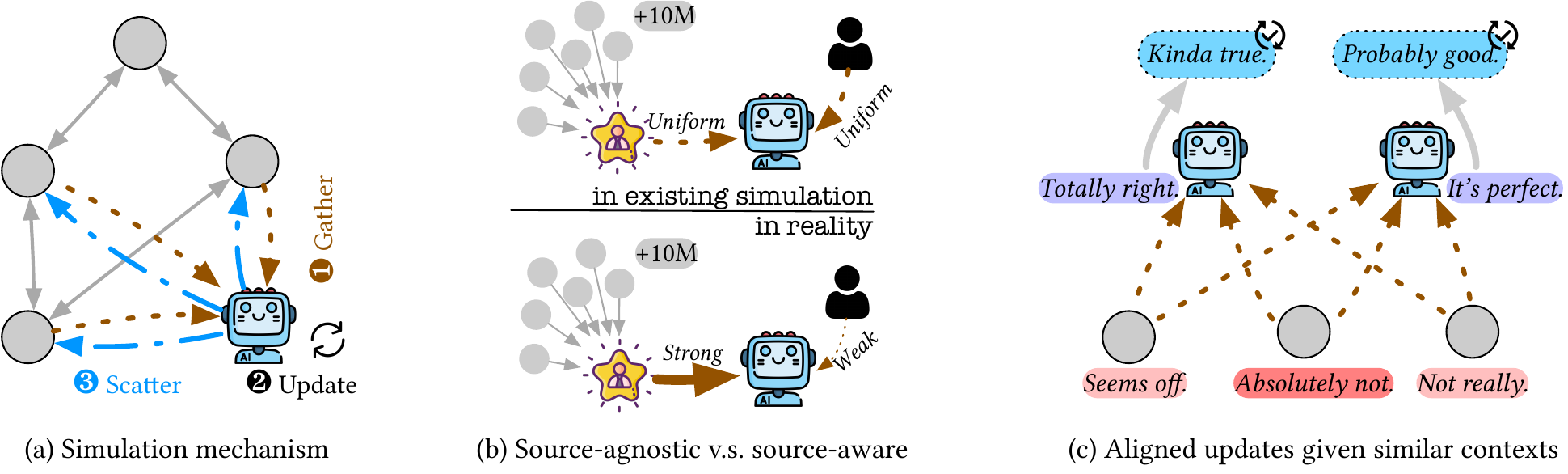}}
    \vspace{-0.4cm}
    \caption{
      Agent behaviors and interaction patterns are coupled with network structure. 
    }
    \label{fig:intro}
  \end{center}
  \vspace{-0.7cm}
\end{figure*}

To instantiate such micro-to-macro simulation, traditional methods model interactions through manually specified or learned update rules over social networks~\cite{1990Friedkin, HegsenKrause2002BCM}.
These rule-based abstractions capture behavior diffusion and opinion dynamics, but real-world social interactions are often contextual and heterogeneous, making fixed update rules difficult to generalize across scenarios.
Recent advances in large language models (LLMs) offer a more flexible alternative, as their reasoning and comprehension capabilities enable agents to mimic human-like cognition and communication~\cite{2023GenerativeAgents}.
Accordingly, LLM-driven social simulation has emerged as a promising paradigm~\cite{yu2025researchtown, 2025ECS, yang2025oasis}, where LLM agents represent individuals and interact iteratively to produce system-level dynamics.
This process resembles iterative message passing over a social network: at each interaction step, an agent \textit{gathers} information from its neighbors, \textit{updates} its internal state with an LLM, and then \textit{scatters} generated messages back to its neighbors, as illustrated in Figure~\ref{fig:intro}(a).
This gather-update-scatter paradigm turns simulation into a graph-structured and stateful LLM workload.

This workload immediately raises a scalability challenge: under conventional execution, each agent is updated independently in each round, causing LLM calls to grow with both simulation nodes and rounds.
Designing scalable systems for data-intensive applications has long been a central focus of the data management community.
Prior work has proposed many systems for data processing~\cite{GraphScope2021Fan, SimDB2022Rod}, semi-structured and unstructured data analysis~\cite{ADBV2020wei, DocDB2025Li}, and machine learning workloads~\cite{jellybean2022Wu,SystemML2016}.
However, comparatively little attention has been paid to scalable system support for social simulation.
In this work, we aim to design a scalable execution system for LLM-driven social simulation.

A closer look at the social graph suggests that this execution challenge is not purely a matter of scale.
Social networks shape not only which agents interact, but also how their interactions unfold, giving rise to topology-induced regularities in behavioral dynamics.
Neighbor influence is inherently asymmetric, where high-profile public figures may exert substantially greater impact than ordinary users for certain receivers (Figure~\ref{fig:intro}(b)).
Meanwhile, individuals with similar structural roles often receive comparable neighborhood exposure and exhibit aligned opinion dynamics due to shared interaction patterns (Figure~\ref{fig:intro}(c)).
Yet current LLM simulators rarely exploit such structural signals at the execution layer, leading to flattened receiver-dependent influence in context construction and redundant inference across compatible agent updates.
This raises a fundamental question: \textit{Beyond serving as a communication scaffold, can network topology act as an execution prior for scalable LLM social simulation?}

Motivated by this question, we propose \method{}, a topology-aware execution layer for LLM-driven social simulation.
Rather than replacing the social logic of existing simulators, \method{} serves as a plug-in layer, introducing topology-derived execution priors.
It uses graph structure, evolving states, and neighbor messages to determine how LLM updates should be materialized and scheduled.
\method{} addresses this through three designs:

\textbf{1) Graph-Structured Workload Formulation.}
We formulate LLM-driven social simulation as repeated gather-update-scatter execution over a graph.
Each agent update assembles local state, profile, memory, and neighbor messages into an LLM request.
Under this formulation, \method{} focuses on the cost-dominant update stage, where it constructs receiver-specific neighbor exposure and coordinates redundant LLM updates for efficiency.

\textbf{2) Topology-Aware Influence Materializer.}
Topology first determines how each receiver should observe its neighborhood.
Agents may share overlapping neighbors, but the influence of each source is receiver-dependent rather than uniform across all agents.
\method{} therefore derives annotation-free, receiver-conditioned influence profiles from the network and uses them to materialize a neighborhood view for the LLM, preserving high-priority messages while compressing or pruning low-influence ones.

\textbf{3) Topology-Aware Update Coordination.}
Topology also informs which agent updates can be coordinated.
In gather-update-scatter execution, agents receiving structurally and semantically similar neighborhood exposure often follow similar update regimes when their current states are close.
\method{} captures these factors with an update signature.
Agents with similar update signatures are grouped into bounded-diameter coordination units, where \method{} executes a representative LLM update instead of invoking the LLM independently for every agent, while keeping the induced one-step update deviation controlled.

Together, the influence materializer and update coordination convert network topology into an annotation-free execution prior for both context construction and LLM execution.
We evaluate \method{} across diverse networked social simulation settings to validate its generality and effectiveness as an execution layer.
Experiments show that \method{} preserves trajectory-level simulation fidelity while reducing token usage by 40--90\% across published simulators, with scale studies confirming consistent resource savings up to 10,000 agents. 
Studies on real-world trajectories further show that the receiver-dependent influence materializer improves alignment with observed social dynamics.

\section{Background}

\subsection{Graph-Based Simulation Formulation} 
From a graph-theoretic perspective, social simulation can be formulated as state evolution over a network.
A social network is represented as a graph \(G=(V,E)\), where each node \(v_i \in V\) corresponds to an agent and each directed edge \((v_j,v_i)\in E\) denotes a channel through which information from \(v_j\) can reach \(v_i\).
At timestep \(t\), agent \(v_i\) maintains a state \(x_i^t\) that encodes its opinion, profile, and memory.
The system evolves through iterative updates by aggregating information from its neighbors:
\begin{equation}
x_i^{t+1} = \mathcal{U}\big( \ x_i^t, \  \{x_j^t : v_j \in \mathcal{N}(v_i)\} \ \big),
\end{equation}
where \(\mathcal{N}(v_i)=\{v_j \mid (v_j,v_i)\in E\}\) and \(\mathcal{U}\) is a general update operator.
This formulation covers both classical opinion dynamics and recent LLM-driven agent simulations.

The key difference is how \(\mathcal{U}\) is instantiated.
Classical models define \(\mathcal{U}\) with hand-crafted rules or numerical update schemes, such as averaging, thresholding, or probabilistic transitions~\cite{1990Friedkin, HegsenKrause2002BCM}.
Since these updates are lightweight, topology mainly shapes the dynamics rather than cost.
LLM-based simulators instead realize \(\mathcal{U}\) through natural language interaction~\cite{yu2025researchtown}. 
Each update constructs an LLM request from the agent's state, memory, and neighbor messages, and then updates the agent state from the response.
This language operator gives LLM-driven social simulation its expressive power while keeping execution cost tied to the graph structure.
Under naive execution, a \(T\)-step simulation over \(|V|\) agents requires \(O(T|V|)\) LLM calls, with cost dominated by repeated prompt-response tokens.
Therefore, cost depends not only on the LLM, but also on how graph-coupled updates are executed.

\begin{figure}[!t]
    \centering
    \includegraphics[width=0.97\linewidth]{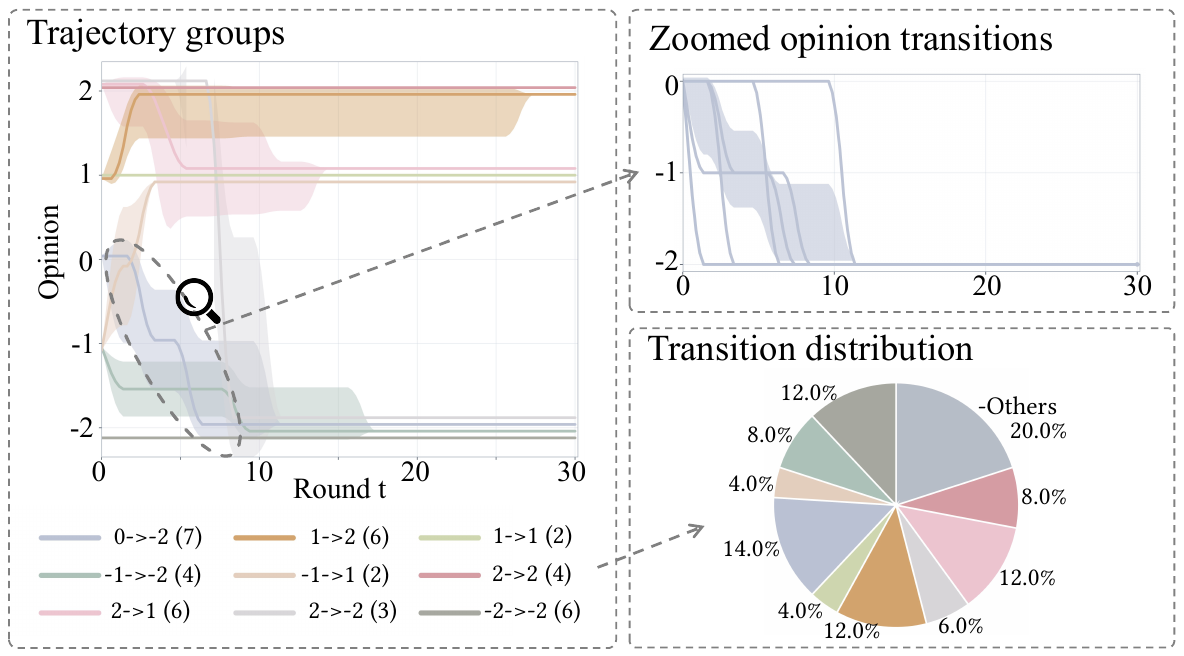}
    \vspace{-0.3cm}
    \caption{Consistency analysis of opinion update trajectory.   } 
    \label{fig:background}
    \vspace{-0.6cm}
\end{figure}

\begin{figure*}[]
  \begin{center}
    \centerline{\includegraphics[width= 0.96\textwidth]{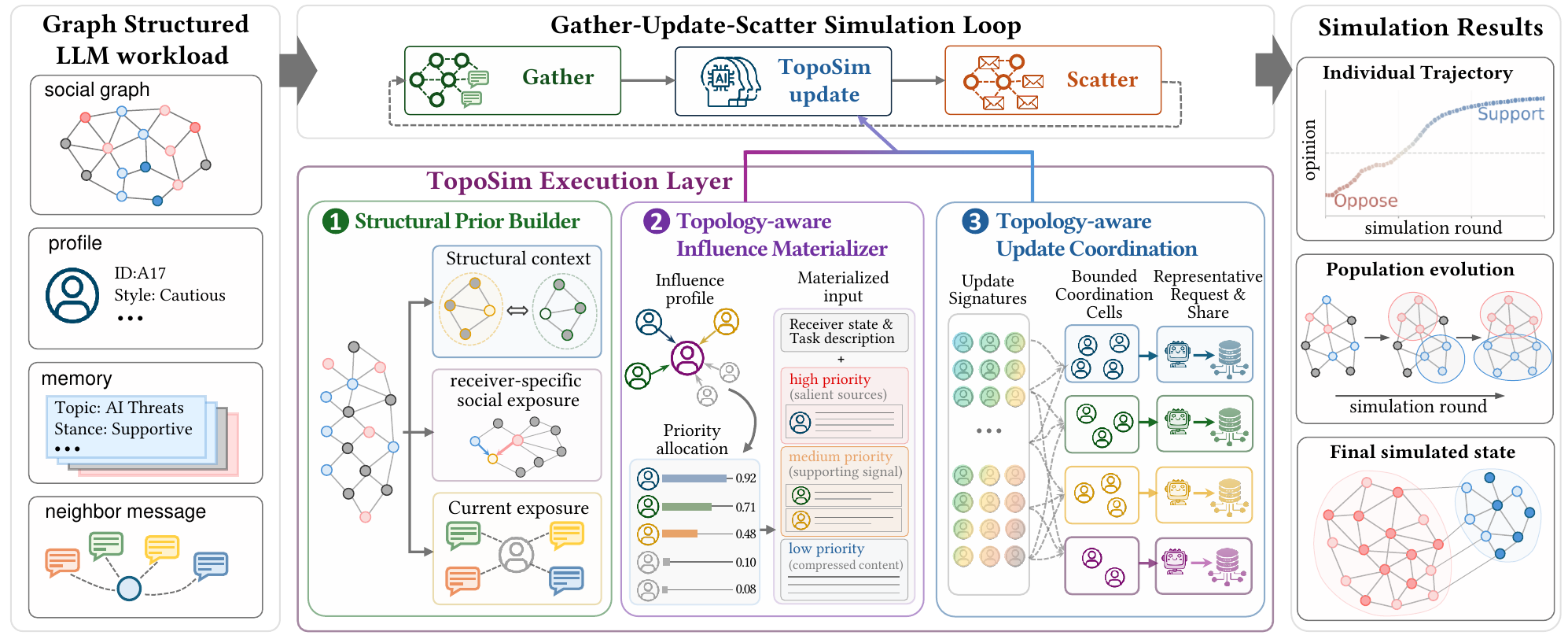}}
    \vspace{-0.3cm}
    \caption{
      Topology-aware execution flow of \method{} for graph-structured LLM simulation.
    }
    \label{fig:method}
  \end{center}
  \vspace{-0.7cm}
\end{figure*}

\subsection{Redundancy in LLM-based Simulation}
Although existing LLM simulators update agents independently, many trajectories may remain aligned because agents receive similar graph-dependent contexts and evolve under similar interaction patterns, suggesting potential execution redundancy.
We examine this issue through a 50-agent EchoChamberSim simulation (ECS-50)~\cite{2025ECS}, where each agent is still updated independently by the LLM.
Figure~\ref{fig:background} shows the trajectory alignment observed in this simulation.
We group agents by opinion trajectories and summarize each group using its median trend and shaded within-group variance bands. Labels of the form \( a \to b \; (n)\) indicate trajectories from initial opinion \(a\) to final opinion \(b\), where \(n\) is the group size.

The grouped trajectories reveal a clear pattern of alignment: the agents with similar initial and final opinions often follow similar intermediate updates rather than fluctuating independently.
A closer look at a representative group reveals consistently low within-group variation across timesteps.
The pie chart shows the distribution of agents across trajectory groups, with 9 groups accounting for about 80\% of all agents.
This observation suggests that redundant LLM updates are not purely incidental; a substantial portion of independently executed agent updates falls into closely related update regimes.
It raises a natural execution question: can such redundancy be identified and reduced without introducing additional supervision, rather than invoking the LLM independently for each agent?
Since the social graph already governs who observes whom and how messages enter each update, \method{} explores topology as an annotation-free execution prior for compressing redundant updates and preserving receiver-specific social exposure.

\section{Method}

In this section, we present \method{}, a topology-aware execution layer for LLM-driven social simulation.
\method{} augments existing gather-update-scatter simulators with an execution layer that uses network topology as execution prior.
We first introduce the overall execution pipeline in Section~3.1, then describe receiver-dependent influence materializer and topology-aware update coordination in Sections~3.2 and~3.3, respectively.
Finally, we analyze the approximation induced by coordination in Section~3.4.

\subsection{\method{} Overview and Execution Model}
We formulate LLM-driven social simulation as a graph-structured, stateful LLM workload.
At each timestep over a social graph, agents execute a \textbf{gather-update-scatter} procedure: they gather local context from their neighborhoods, apply an LLM-based operator to update their states, and scatter updated states or messages back into the graph for iterative interactions.
Conventional execution schedules this procedure independently for every agent at every round.
Under this schedule, each update is issued as a separate LLM request, even though its input depends on the graph-neighborhood context and its output affects later state and message propagation.
This creates the central execution mismatch: the simulation is organized by graph, but LLM execution remains largely topology-agnostic.

\method{} addresses this mismatch by treating topology not only as a simulation constraint, but also as an execution prior.
This prior is derived from the graph and instantiated with evolving states, rather than manually assigned social roles or external influence labels.
It enters execution through two complementary mechanisms.
First, the influence materializer uses structural asymmetry to construct neighborhood views, because the same source may carry different, rather than uniform, influences for different receivers.
Second, update coordination uses structural similarity, combined with realized exposure and current agent states, to identify update requests that can share LLM execution.
Together, these mechanisms make topology actionable at the execution layer, shaping both the context for LLM and LLM invocation.

Figure~\ref{fig:method} illustrates the execution flow of \method{}.
Given the input graph and initial states, \method{} first constructs structural priors, including stable structural context and receiver-dependent influence profiles.
The materializer converts these profiles into prioritized neighborhood views, producing receiver-specific LLM update requests.
The execution planner then combines structural context, current states, and realized exposure to form coordinated inference cells and select representative requests for execution.
After the representative LLM updates are generated, their results are projected back to the agents and returned to the backend simulator for the next round.
Thus, the backend simulator still defines the original social update logic, while \method{} only controls how neighbor information is presented and how LLM updates are scheduled.

This framework turns the social graph from a passive communication scaffold into an active execution prior for scalable LLM social simulation.
Sections 3.2--3.4 detail these components and analyze the approximation induced by coordination.

\subsection{Topology-Aware Influence Materializer}

The first component of \method{} specifies how neighborhood information is exposed to the LLM for each receiver.
In gather-update-scatter simulators, the LLM does not operate on the graph directly, but on textual context aggregated from neighbor messages.
If all messages are presented as a uniformly weighted list, the execution layer flattens the heterogeneous exposure patterns induced by the social graph.
This is especially problematic in directed networks, where the same source may be prominent to one receiver but marginal to another.
\method{} therefore uses a receiver-dependent influence materializer to convert topology-induced asymmetry into the neighbor context that the LLM can consume.

We formalize this topology-to-context conversion as a budget-constrained materializer.
For a receiver \(v_i\), the materializer constructs an influence-aware prompt block:
\begin{equation}
\mathcal{P}_i^t =
\mathcal{M}
\left(
\{(\pi_j^i,\Phi(x_j^t)) : v_j\in\mathcal{N}(i)\}; B
\right),
\end{equation}
where \(\Phi(x_j^t)\) extracts the communicative message from the current state of \(v_j\), \(\pi_j^i\) denotes the structural influence of \(v_j\) with respect to receiver \(v_i\), \(B\) is the prompt budget, and \(\mathcal{M}\) is the materializer operator.
The set passed to \(\mathcal{M}\) contains the neighbor messages available to \(v_i\) at timestep \(t\), each paired with its receiver-specific influence weight.
Given this weighted message set, \(\mathcal{M}\) translates topology-induced influence into an LLM-readable textual layout.
It determines the ordering, grouping, and compression level of neighbor messages, so that structurally important sources receive more prominent exposure while insignificant signals remain available in a compact form.

We instantiate \(\pi^i\) using Personalized PageRank on the graph:
\begin{equation}
\pi^i = (1-\alpha)e_i + \alpha W\pi^i,
\end{equation}
where \(e_i\) is the personalization vector of receiver \(v_i\), \(W\) is the normalized transition matrix, and \(\alpha\) is the damping factor.
Following the graph convention in Section~2, an edge \((v_j,v_i)\) indicates that information from \(v_j\) can reach \(v_i\), and \(W\) is normalized along this information-flow structure.
Therefore, \(\pi_j^i\) is not a global importance, but an accessibility score measuring how strongly source \(v_j\) appears in the structural view of receiver \(v_i\).
The personalization vector makes \(\pi^i\) receiver-specific, while propagation through \(W\) incorporates both direct and multi-hop reachability in the graph.

Given this profile, \(\mathcal{M}\) materializes neighbor messages by influence priority rather than numerical weights.
Messages from high-\(\pi_j^i\) sources are placed earlier and preserved with greater detail, while lower-priority ones are summarized or truncated under context budget.
This converts topology-induced asymmetry into an LLM-readable context organized by structural priority.
In this way, the influence profile shapes the receiver's observed social context without manually assigned roles or explicit labels.
The materialized prompt block \(\mathcal{P}_i^t\) serves as the receiver-specific neighborhood context for the LLM update.
Its influence-aware construction also yields the exposure signal used later by the execution planner.

\subsection{Topology-Aware Update Coordination}
While receiver-specific neighborhood context has been materialized, naive execution would still invoke the LLM once for every agent at every timestep.
However, the redundancy observed in Section~2.2 suggests that many independently executed LLM updates may fall into closely related update regimes.
The goal of update coordination is therefore to identify, before LLM invocation at timestep \(t\), which graph-coupled update requests can share a representative execution while keeping the induced update deviation controlled.
We formulate this objective as a radius-constrained covering problem over projected update distributions.

Let \(q_i^t\) denote the complete update request of agent \(v_i\) at timestep \(t\).
After prompt construction, LLM generation, and opinion projection, this request induces a projected update distribution
\begin{equation}
P_i^t = \bar{\mathcal{U}}(\cdot \mid q_i^t),
\end{equation}
where \(\bar{\mathcal{U}}\) denotes the simulator-level projected update operator.
Given an admissible deviation radius \(\eta\), the ideal coordination objective is to cover all requests with a minimal set of representatives:
\begin{equation}
\min\nolimits_{\mathcal{C}_t,\{r_C\}} |\mathcal{C}_t|
\quad \mathrm{s.t.}
\quad D(P_i^t,P_{r_C}^t)\le \eta,
\quad \forall C\in\mathcal{C}_t,\ v_i\in C .
\end{equation}
The covering objective provides a template for the planner: the radius specifies which requests may share a representative, the objective favors fewer representatives, and the representative should stay close to all requests it covers.
The challenge is that the true distance \(D(P_i^t, P_j^t)\) is unavailable before LLM execution.
\method{} thus keeps the covering structure, but replaces the ideal post-execution distance with an observable topology-conditioned surrogate.

To construct such a surrogate for this distance, \method{} records the observable factors that shape an update request before the LLM is called.
These factors separate three sources of update similarity.
First, an agent's structural position determines its persistent graph context and the type of information it tends to receive and propagate.
Second, the agent's stance and resistance to opinion change characterize its intrinsic response condition.
Third, the receiver-specific exposure summarizes the neighbor messages materialized at timestep \(t\).
We collect these factors into an update signature:
\begin{equation}
\xi_i^t = \big(\mu_i,\ o_i^t,\ s_i,\ e_i^t\big),
\end{equation}
where \(\mu_i\) encodes the structural position of \(v_i\), \(o_i^t\) denotes the current stance, \(s_i\) captures resistance to opinion change, and \(e_i^t\) summarizes the current receiver-specific exposure.
Unlike the persistent structural term \(\mu_i\), the exposure term \(e_i^t\) varies with the current neighbor states:
\begin{equation}
e_i^t = \sum\nolimits_{j\in\mathcal{N}(i)} \pi_j^i \mathbf{e}\big(\Phi(x_j^t)\big).
\end{equation}
Here \(\Phi(x_j^t)\) maps a neighbor's state to an opinion label, and \(\mathbf{e}(\cdot)\) embeds the label into a vector representation.
\(\pi_j^i\) is the receiver-dependent influence weight of \(v_j\) for \(v_i\), defined by the materializer in Section~3.2.
The signature is not intended to reproduce the full prompt; it records the observable factors most directly related to whether two requests can update similarly.

Given these signatures, \method{} compares update requests from two agents through their normalized execution distance:
\[
d_{\mathrm{exec}}^t(i,j)
=
\max
\left\{
\frac{d_{\mathrm{str}}(\mu_i,\mu_j)}{\epsilon_\mu},
\frac{|o_i^t-o_j^t|}{\epsilon_o},
\frac{|s_i-s_j|}{\epsilon_s},
\frac{d_{\mathrm{exp}}(e_i^t,e_j^t)}{\epsilon_e}
\right\}.
\]
Here, \(d_{\mathrm{str}}\) measures structural dissimilarity, and \(d_{\mathrm{exp}}\) compares neighborhood exposure vectors.
The denominators instantiate the admissible radius in the signature space: \(d_{\mathrm{exec}}^t(i,j)\le 1\) holds only when the requests remain within the allowed range along every update-relevant factor. 
This criterion supports staged planning, where the structural component $\mu$ first filters agents with compatible graph contexts before timestep-specific exposure is refined.

The planner applies this covering principle in surrogate space.
A coordination unit is built as a bounded-diameter execution cell:
\begin{equation}
\operatorname{diam}_{\mathrm{exec}}^t(C)
=
\max\nolimits_{i,j\in C} \ d_{\mathrm{exec}}^t(i,j),
\quad
\operatorname{diam}_{\mathrm{exec}}^t(C)\le 1.
\end{equation}
All requests inside the same cell stay within the specified radius and are therefore planned for shared execution.
Using observable signatures, the planner builds a candidate graph connecting request pairs with \(d_{\mathrm{exec}}^t(i,j)\le 1\), and greedily forms cells while maintaining the bounded-diameter constraint.
Each feasible cell \(C\) corresponds to one LLM invocation, with representative request
\begin{equation}
r_C = \arg\min_{i\in C} \max_{j\in C} d_{\mathrm{exec}}^t(i,j).
\end{equation}
This choice minimizes the worst-case execution distance from the representative to other requests in the cell.
The simulator then invokes the LLM once for \(r_C\) and reuses the projected update within \(C\).
The perturbation thresholds control the efficiency--fidelity trade-off: tighter thresholds create finer cells and preserve individualized updates, while looser ones enable stronger compression.
We then analyze how this bounded-diameter construction controls the one-step approximation error of coordinated execution.

\subsection{Approximation of Coordinated Execution} 

Coordinated execution replaces multiple per-agent LLM calls with one representative call, thereby introducing approximation.
We analyze this approximation at the level of projected opinion states rather than raw generated text.
This matches the simulator interface: after each LLM update, the generated response is mapped by \(\Phi(\cdot)\) to an opinion category or low-dimensional opinion state, which is then used in subsequent social dynamics analysis.
Let
\begin{equation}
Y_i^{t+1}=\Phi(x_i^{t+1}), \quad
Y_i^{t+1}\sim P_i^t
=
\bar{\mathcal{U}}(\cdot\mid \xi_i^t),
\end{equation}
where \(\xi_i^t=(\mu_i,o_i^t,s_i,e_i^t)\) is the update signature in Section~3.3, and \(\bar{\mathcal{U}}\) denotes the simulator process that combines prompt construction, LLM generation, and opinion projection.
To link the planner's surrogate distance to the actual projected deviation, we require the following local stability condition at the projection level.

\begin{assumption}[Local stability of projected updates]
For update signatures \(\xi\) and \(\xi'\), the projected update distributions satisfy
\begin{equation}
D\left(
\bar{\mathcal{U}}(\cdot\mid \xi),
\bar{\mathcal{U}}(\cdot\mid \xi')
\right)
\le
\omega\left(d_{\mathrm{exec}}(\xi,\xi')\right),
\end{equation}
where \(D(\cdot,\cdot)\) is a divergence between projected opinion distributions, \(d_{\mathrm{exec}}\) is the signature-level execution distance corresponding to Section~3.3, and \(\omega(\cdot)\) is non-decreasing stability modulus with \(\omega(0)=0\).
\end{assumption}

This assumption concerns the simulator-facing projected states rather than raw language outputs.
It formalizes the execution-level regularity that nearby update signatures induce nearby projected dynamics.
Section~4 empirically supports this regularity through the within-cell full-execution variance in Figure~\ref{fig:coord_within}(b), showing \(d_{\mathrm{exec}}\)-grouped agents remain similar under independent execution.

\begin{proposition}[One-step coordination error]
For any coordination cell \(C\) with representative \(r_C\), the projected update deviation of each agent \(v_i\in C\) satisfies
\begin{equation}
D(P_i^t,P_{r_C}^t)
\le
\omega\left(
\operatorname{diam}_{\mathrm{exec}}^t(C)
\right).   
\end{equation}
Thus, when \(\operatorname{diam}_{\mathrm{exec}}^t(C)\le 1\), we have \( \; D(P_i^t,P_{r_C}^t)\le \omega(1) . \)
\end{proposition}

The result follows from the definition of execution diameter:
\(d_{\mathrm{exec}}^t(v_i,r_C)\le \operatorname{diam}_{\mathrm{exec}}^t(C)\), together with Assumption~1.
Therefore, bounded-diameter cells provide an executable surrogate for the ideal radius-constrained covering objective in Section~3.3.
When the perturbation thresholds are calibrated such that \(\omega(1)\le\eta\), each representative update satisfies the admissible one-step deviation radius of the ideal formulation.

The maximum form of \(d_{\mathrm{exec}}\) makes the bound depend on the largest normalized discrepancy across signature factors. 
Thus, a large discrepancy in structure, stance, resistance, or receiver-specific exposure prevents two requests from sharing a cell, even when they are close otherwise.
This connects bounded-diameter cells to the local stability condition and makes the efficiency--fidelity trade-off explicit.
Smaller cells reduce \(\operatorname{diam}_{\mathrm{exec}}^t(C)\)  and tighten the bound, while larger cells save more LLM calls with a looser bound.

\section{Experiment}

\begin{figure*}[!t]
    \centering
    \includegraphics[width=0.99\textwidth]{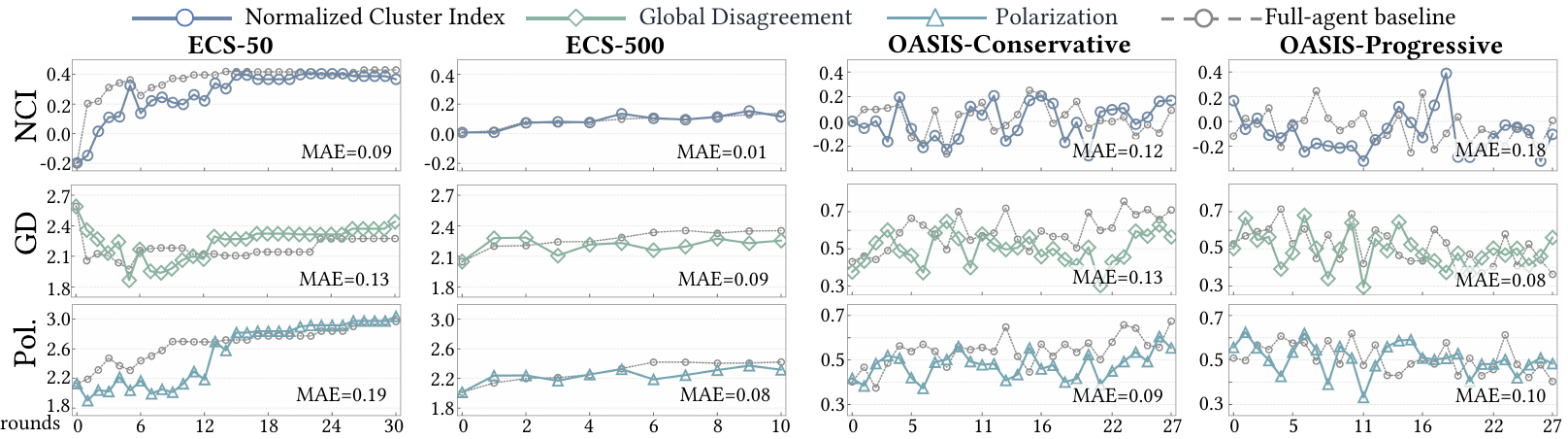}
    \vspace{-0.4cm}
    \caption{Macro-level fidelity of coordinated execution on ECS and OASIS.}
    \label{fig:coord_macro}
    \vspace{-0.4cm}
\end{figure*}

We evaluate \method{} as a plug-in execution layer for graph-coupled LLM social simulation on simulator backends.
The evaluation asks whether topology-derived priors reduce LLM use, preserve fidelity, expose cost--fidelity trade-offs, and generalize across settings.
We organize experiments around four questions:
\circledchar{$Q_1$} Can topology-aware coordination reduce LLM execution while preserving trajectory fidelity?
\circledchar{$Q_2$} How do topology-derived priors trade off execution cost and fidelity?
\circledchar{$Q_3$} Which components drive the gains, and do they improve real-trajectory alignment?
\circledchar{$Q_4$} Do the benefits persist across simulator workloads, LLMs, and graph scales?

\subsection{Experimental Setup}

\textbf{Execution setting.}
All experiments run on an Ubuntu 24.04 server with Xeon Gold 6330 CPUs, 1 TB RAM, and an RTX A5000 GPU.
Unless otherwise specified, we use \texttt{gpt-4o-mini} with temperature 0.7, following each simulator's default configuration.
We also evaluate two alternative LLM backbones for cross-LLM robustness.
Topology signals use Personalized PageRank~\cite{jeh2003personalpagerank} with damping factor 0.85 and 128-dimensional \textsc{struc2vec}~\cite{Ribeiro2017struc2vec} embeddings.

\textbf{Workloads.}
We evaluate \method{} on three published simulator backends, forming a workload matrix with complementary validation roles, as summarized in Table~\ref{sample_table}.
\textsc{EchoChamberSim (ECS)}~\cite{2025ECS} provides controlled opinion-polarization workloads with flexible scales and full-agent reference trajectories, enabling evaluation of execution compression.
\textsc{OASIS}~\cite{yang2025oasis} uses real user profiles in a social media simulator, enabling coordination tests with richer agent attributes and simulated platform interactions.
\textsc{FDE-LLM}~\cite{Yao2025FDEllm} is grounded in real discussions with observed trajectories, offering a scarce real-world reference for validating trajectory alignment.

\vspace{-0.25cm}
\begin{table}[!h]
  \caption{Workload roles and statistics.}
  \vspace{-0.4cm}
  \small
  \label{sample_table}
  \centering
  \begin{sc}
    \begin{tabular}{lcccl}
    \toprule
    Workload  & \# Nodes  & \# Edges & \# Rounds & Role \\
    \midrule
    ECS-50    & 50  & 108   & 30 & Controlled ref. \\
    ECS-500   & 500 & 1,753 & 10 & Scale ref. \\
    OASIS     & 196 & 1,279 & 80 & Real-profile sim \\
    FDE-LLM   & 206 & 1,503 & 18 & Observed traj. \\
    \bottomrule
    \end{tabular}
  \end{sc}
\vspace{-0.25cm}
\end{table}

\textbf{Metrics.}
For update coordination, full-agent execution under the same original social update process provides uncompressed reference trajectories.
For materialization, we compare with uniform exposure on FDE-LLM, where observed trajectories provide real-world references.
Fidelity is measured by MAE to the corresponding reference: polarization (Pol.), global disagreement (GD), and normalized clustering index (NCI) for macro dynamics, and per-agent opinion trajectories for micro-level fidelity.
For execution cost, we report LLM calls and tokens relative to full-agent execution, since \method{} reduces LLM workload rather than backend scheduling.
Calls reflect the inference units before platform-specific scheduling, while tokens determine both monetary cost and provider-side serving load. 
Together, they isolate execution-layer savings better than wall-clock latency.

\subsection{Execution Efficiency via Coordination}
To answer \circledchar{$Q_1$}, we evaluate whether \method{} reduces LLM execution while preserving trajectory fidelity.
We focus on ECS and OASIS, where full-agent runs under the same original social update process provide direct reference trajectories, while FDE-LLM is reserved for further real-trajectory validation.

\textbf{Macro fidelity.}
We first evaluate whether coordination preserves macro-level dynamics.
Figure~\ref{fig:coord_macro} compares full-agent and coordinated execution on ECS and OASIS using Pol., GD, and NCI trajectories.
Across various workload settings, coordinated execution closely follows the full-agent trajectories, with only small transient deviations at early timesteps.
Across the reported metric trajectories, most MAE values are around 0.1, which is small relative to the observed ranges of Pol. and GD, and remains modest even for the narrower NCI range.
These results show that representative updates preserve macro-level simulation trends under substantial execution sharing, without biasing aggregate dynamics.

\begin{figure}[!t]
    \centering
    \includegraphics[width=0.94\linewidth]{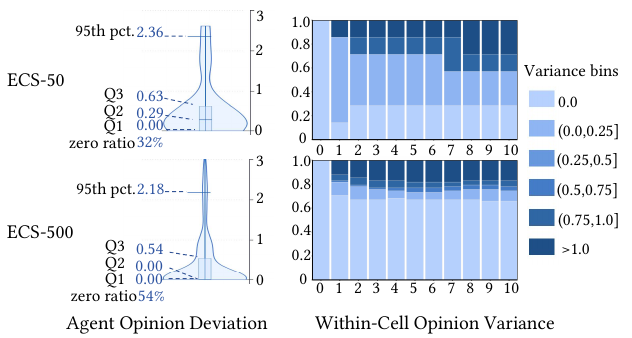}
    \vspace{-0.4cm}
    \caption{Micro-level fidelity of coordination.}
    \label{fig:coord_within}
    \vspace{-0.5cm}
\end{figure}

\begin{figure}[!t]
    \centering
    \includegraphics[width=0.66\linewidth]{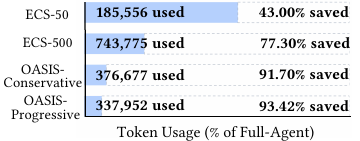}
    \vspace{-0.4cm}
    \caption{Token cost of coordination across workloads.}
    \label{fig:coord_cost}
    \vspace{-0.6cm}
\end{figure}

\textbf{Micro fidelity.}
We use micro-level fidelity as an empirical diagnostic for the local stability condition behind bounded-diameter cells.
Figure~\ref{fig:coord_within}(a) reports per-agent opinion-trajectory MAE between full-agent and coordinated execution.
The differences are concentrated near zero with a right-skewed tail. Exact matches account for 32\% of agents on ECS-50 and 54\% on ECS-500, while the upper quartiles remain below 0.65.
Thus, most agents stay close to their full-agent trajectories.
Figure~\ref{fig:coord_within}(b) measures within-cell variance under full-agent execution, testing whether cells built from pre-execution signatures correspond to post-hoc trajectory similarity.
In ECS-500, over 80\% of cells have variance below 0.25, corresponding to a standard deviation below 0.5, roughly half an opinion level.
Together, these results support bounded-diameter planning: pre-execution cells identify agents with locally similar trajectories.

\textbf{Execution efficiency.}
Figure~\ref{fig:coord_cost} reports token usage relative to full-agent execution, treating them as deployment-facing resource units directly affected by the execution layer.
Coordination consistently reduces token usage across workloads, with token savings ranging from 43.0\% to 93.4\%.
Savings further increase on larger ECS graphs, suggesting that scale exposes more reusable update regimes.
OASIS has the largest token reduction because its profile-rich agent contexts make each redundant prompt more expensive, so sharing updates saves more.
Overall, coordination turns pre-execution similarity into substantial resource reduction while keeping macro- and micro-level trajectory errors small.

\begin{figure}[!t]
    \centering
    \includegraphics[width=0.76\linewidth]{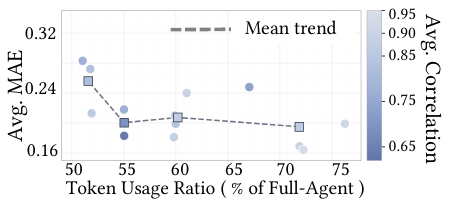}
    \vspace{-0.35cm}
    \caption{Cost-fidelity operating frontier.}
    \label{fig:trade_off}
    \vspace{-0.4cm}
\end{figure}

\subsection{Cost--Fidelity Trade-off}
To answer \circledchar{$Q_2$}, we examine how coordination thresholds expose operating points between execution cost and trajectory fidelity.
This analysis follows from the bounded-diameter design: tighter thresholds restrict sharing to more similar update requests, whereas looser thresholds allow broader sharing and stronger compression.

\textbf{Operating trade-off.}
We treat coordination thresholds as operational controls on ECS-50.
Figure~\ref{fig:trade_off} plots full-agent normalized token ratio against opinion-trajectory MAE, with color indicating average trajectory correlation.
These thresholds define the admissible execution radius: tighter settings form smaller cells and spend more tokens, whereas looser settings permit broader sharing.
The figure shows how one-step coordination tolerance translates into multi-step trajectory fidelity.
Instead of a single fixed setting, \method{} exposes a family of deployment choices under different resource budgets.
The fidelity-oriented setting achieves the lowest MAE at higher token usage, while a balanced setting around 60\% token ratio retains both low MAE and high correlation.
Token usage can drop further to about 52\% with modest MAE, but some low-token settings show clear correlation loss.
This indicates that coordination is tunable, but overly aggressive sharing eventually weakens temporal alignment.

\subsection{Component Analysis and Real-World}
To answer \circledchar{$Q_3$}, we identify the components behind the gains and test alignment with real social trajectories.
Because observed trajectories are scarce, we use the Weibo dataset from FDE-LLM, where Weibo is a large Chinese microblogging platform.
Unlike ECS and OASIS, which provide simulator-reference comparisons, this dataset contains observed attitude trajectories from real discussions.

\begin{figure}[!t]
    \centering
    \includegraphics[width=0.99\linewidth]{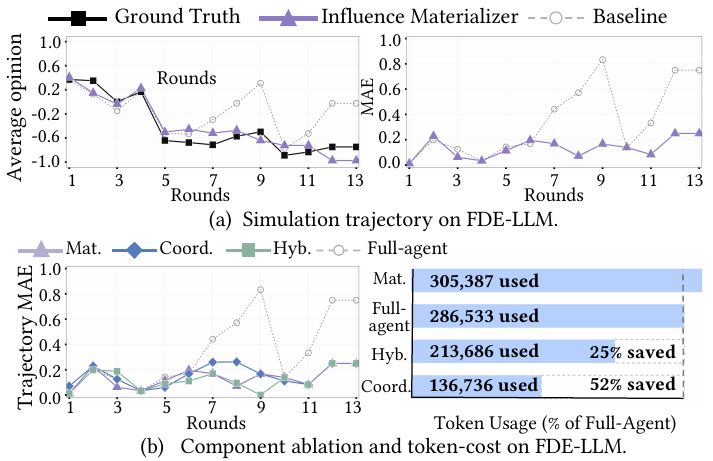}
    \vspace{-0.35cm}
    \caption{Real-trajectory alignment and component ablation on FDE-LLM.}
    \label{fig:ablation_fidelity}
    \vspace{-0.4cm}
\end{figure}

\textbf{Real-trajectory alignment.}
Figure~\ref{fig:ablation_fidelity}(a) compares the ground truth (GT), the uniform-exposure baseline, and the materialized variant (\textit{Mat.}).
The baseline gradually drifts from the observed trajectory, whereas \textit{Mat.} better follows the direction and magnitude of the ground-truth trend.
The MAE curve further shows that \textit{Mat.} maintains higher alignment with GT over time.
This indicates that materialization preserves asymmetric influence patterns common in real social platforms and improves trajectory fidelity.

\begin{table}[!t]
  \caption{Coordination signal ablation.}
  \vspace{-0.4cm}
  \small
  \label{CorrelationAblation}
  \centering
  \begin{sc}
    \begin{tabular}{lccccc}
    \toprule
    Signal  
    & \makecell{NCI \(\downarrow\)}
    & \makecell{GD \(\downarrow\)}
    & \makecell{Pol \(\downarrow\)}
    & \makecell{Avg. MAE \(\downarrow\)}
    & \makecell{Avg. \(r\) \(\uparrow\)} \\
    \midrule
    Random   & 0.28 & 0.21 & 0.53 & 0.34 & -0.06 \\
    State-only  & \textbf{0.04} & 0.82 & 1.18 & 0.68 & 0.11  \\
    Topo.-only     & 0.27 &0.55 & 1.13 & 0.58 & -0.11  \\
    \(K\)-means  & 0.08 & 0.38 & 0.46 & 0.31 & 0.40 \\
     Ours    & 0.08  & \textbf{0.13}   & \textbf{0.20}& \textbf{0.14} & \textbf{0.78}  \\
    \bottomrule
    \end{tabular}
  \end{sc}
\vspace{-0.5cm}
\end{table}

\textbf{Component ablation on FDE-LLM.}
Figure~\ref{fig:ablation_fidelity}(b) compares full-agent execution, coordination only (\textit{Coord.}), materializer only (\textit{Mat.}), and their hybrid on FDE-LLM, along with token usage.
\textit{Coord.} reduces tokens but later loses alignment, suggesting compression alone may accumulate deviation.
In contrast, \textit{Mat.} better follows the observed trajectory through receiver-specific exposure, but keeps cost close to full-agent execution.
The hybrid combines their strengths: coordination reduces tokens, while materialization recovers trajectory fidelity.
These results show that the two components address complementary aspects of topology-aware execution.

\textbf{Coordination signal ablation.}
Table~\ref{CorrelationAblation} isolates whether coordination needs the full design in Section~3.3.
\method{} combines topology-derived structure with state consistency, and forms bounded-diameter cells by execution distance.
On ECS-50, we compare against state-only and topology-only ablations, k-means over the same features, and random grouping with the same number of coordination units.
We report end-of-run MAE to the full-agent baseline for Pol., NCI, GD, their mean, and average Pearson correlation (Avg-r) to baseline.
\method{} achieves the lowest overall MAE (0.14) and highest Avg-r (0.78).
K-means is the closest alternative, but still has more than twice the overall MAE (0.31) and much lower correlation (0.40).
State-only and topology-only variants fail to preserve macro trajectories consistently, while random grouping performs poorly under the matched token budget.
Together, these results show that effective coordination requires both structural and state consistency, rather than a single signal or generic clustering.

\begin{figure}[!t]
    \centering
    \includegraphics[width=0.99\linewidth]{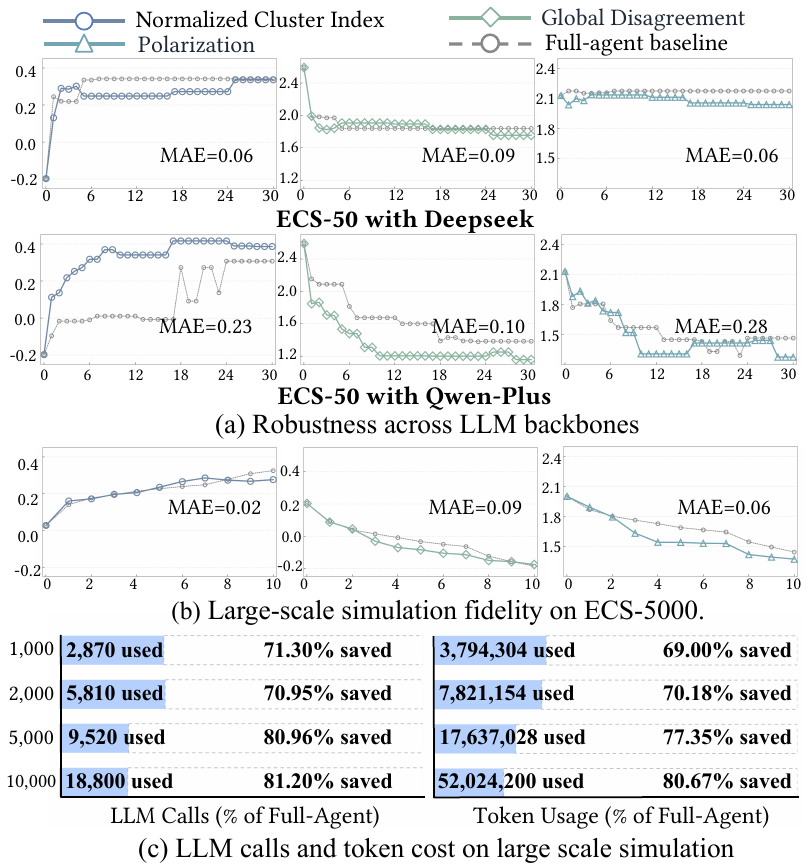}
    \vspace{-0.35cm}
    \caption{Robustness and scalability on ECS.}
    \label{fig:scala_scale}
    \vspace{-0.4cm}
\end{figure}

\subsection{Robustness and Scalability}
To answer \circledchar{$Q_4$}, we examine whether scalability benefits of \method{} persist across LLM backbones and graph scales.
Together with previous cross-workload results, this study evaluates topology-derived execution priors across workloads, models, and workload sizes.

\textbf{Cross-LLM robustness.}
We evaluate coordination on ECS with two additional LLM backbones, \texttt{Qwen-Plus} and \texttt{DeepSeek-V3.2}.
In addition to the cross-workload results in Figure~\ref{fig:coord_macro}, this experiment complements robustness evaluation by testing different models on the ECS-50 workload.
As shown in Figure~\ref{fig:scala_scale}(a), coordinated execution closely tracks full-agent execution for both backbones.
This suggests that the coordination prior is not tied to a single LLM response style, but remains effective across update operators.

\textbf{Scale robustness.}
We then stress coordinated execution on larger ECS graphs to examine whether reusable update regimes persist with scale.
Figure~\ref{fig:scala_scale}(b) shows that coordinated trajectories continue to track full-agent reference trajectories on large graphs.
Figure~\ref{fig:scala_scale}(c) further reports consistent token reduction as the scale increases to 10,000 agents.
These results show that coordinated execution preserves trajectory fidelity at large scale, while its resource savings remain scalable up to 10,000 agents.

\section{Related Work}

\textbf{Social Simulation.}
Social simulation studies how collective social phenomena emerge from individual interactions.
Classical approaches formalize this process through analytical or statistical formulations, where localized influence rules, such as Friedkin--Johnsen opinion updating and bounded-confidence interactions, govern state evolution over social networks~\cite{1990Friedkin, Deffuant2000MixingBA, Kan2023BCM}.
Agent-based models further instantiate this bottom-up view by representing society as collections of autonomous agents interacting in structured environments~\cite{1996Epstein, Watts1998Collective, Barabsi1999EmergenceOS, Bernstein2013agent}. 
Recent LLM-driven simulators extend this paradigm with language-mediated agents that reason over profiles and social context, enabling more expressive simulations than prescribed rules~\cite{2023GenerativeAgents, yu2025researchtown}.
In networked settings, such agents have been used to study collective dynamics, including opinion evolution, emotion propagation, and message diffusion~\cite{gao2023s3, Ferraro2025echo, 2024-opiniondynamics, liu2025fakenews}.
These studies mainly improve the modeling side of social simulation.
In contrast, \method{} focuses on the execution layer, using network topology as an execution prior to improve the efficiency and fidelity of LLM-driven simulation.

\textbf{Structural Influence.}
Social network analysis has long recognized that behavior and influence are shaped by structural position.
Nodes occupying similar structural roles can exhibit related interaction patterns even when their attributes differ~\cite{lorrain1971structural, white1976social}.
This insight has motivated role-oriented graph modeling, from blockmodels~\cite{white1976social} to computational role discovery and structural representation methods such as RolX and struc2vec~\cite{RolX2012, Ribeiro2017struc2vec}.
Meanwhile, centrality and random-walk-based measures, including personalized PageRank, have been widely used to characterize asymmetric prominence and accessibility in networked processes~\cite{page1999pagerank, jeh2003personalpagerank, donnat2018trackingdynamics, aral2012identifying}.
Together, these studies show that topology provides informative cues for networked simulation.
While prior work mainly uses such cues to characterize roles or influence, \method{} leverages them as execution priors for efficient and faithful LLM-driven simulation.

\textbf{Scalable Execution.}
The graph-coupled dynamics above make LLM social simulation a data-intensive execution problem beyond modeling: each step invokes costly LLM updates and maintains evolving agent states.
Data-management research has long turned related data-intensive workloads into optimizable execution problems.
Earlier systems expose this view through in-database analytics~\cite{MadLib2012}, declarative ML compilation~\cite{SystemML2016}, graph-parallel iteration~\cite{GraphLab2012Low}, and database-valued stochastic simulation~\cite{SimSQL2013Cai}.
More recent systems additionally extend this view to large-graph processing and incremental graph updates~\cite{GraphScope2021Fan, Ingress2021Gong}.
Together, they show that data access, scheduling, and state maintenance should be targets for optimization.
Existing LLM social-simulation platforms, by contrast, mainly support scalability through agent engines, orchestration layers, and customizable interfaces~\cite{piao2025agentsociety, yang2025oasis, zhou2025-sotopia}.
\method{} takes a different angle: it treats networked LLM simulation as a topology-aware execution workload, using graph structure to materialize receiver-specific contexts and coordinate redundant LLM updates.

\section{Conclusion}
In this work, we revisit networked LLM social simulation as a graph-structured, stateful LLM workload whose scalability depends on how graph-coupled updates are executed.
We present \method{}, a topology-aware execution layer that uses network topology as an annotation-free execution prior: receiver-dependent influence materializes asymmetric neighbor exposure, while topology-aware coordination shares compatible LLM updates.
Across workloads, LLM backbones, and graph scales, \method{} preserves trajectory fidelity, improves alignment with observed dynamics, and exposes a practical cost--fidelity trade-off.
While the current design focuses on static graphs and execution planning, future work can extend \method{} to evolving networks, incremental topology maintenance, or parallel scheduling.
Overall, \method{} shows that topology can guide not only how social interactions are modeled, but also how graph-structured LLM simulation workloads are executed.


\bibliographystyle{ACM-Reference-Format}
\bibliography{sample}

\end{spacing}

\end{document}